\documentclass{article}

\usepackage{PRIMEarxiv}

\usepackage[utf8]{inputenc}
\usepackage[T1]{fontenc}
\usepackage{hyperref}
\usepackage{url}
\usepackage{booktabs}
\usepackage{amsfonts}
\usepackage{nicefrac}
\usepackage{microtype}
\usepackage{graphicx}
\graphicspath{{figures/}{media/}}

\usepackage{iftex}
\ifPDFTeX\else
  \usepackage{fontspec}
\fi

\usepackage{csquotes}
\usepackage[style=numeric,sorting=none,backend=biber]{biblatex}
\title{TRAIL: A Platform for Configurable Human--AI Teaming Experiments}

\author{
  {\normalfont Mohammad Amin Samadi\thanks{Corresponding author: \texttt{masamadi@uci.edu}} \quad Pedro Martins De~Bastos \quad Jaeyoon Choi} \\
  {\normalfont Spencer JaQuay \quad Seehee Park \quad Nia Nixon} \\[7pt]
  {\normalfont School of Education, University of California, Irvine, Irvine, CA, USA} \\
  {\normalfont\texttt{\{masamadi,\ pedrom4,\ jaeyooc3,\ sjaquay,\ seeheep,\ dowelln\}@uci.edu}}
}

\begin{document}
\maketitle

\begin{abstract}
An AI teammate's design properties (personality, communication style, when it speaks) can shape a team's trust, coordination, and decisions. Studying this rigorously demands infrastructure no existing tool provides: reproducible configuration of an AI teammate embedded in instrumented, real-time collaboration sustained over time. We present the Team Research and AI Integration Lab (TRAIL), a web platform that makes the AI teammate a configurable, reproducible design object, pairing a Big Five persona with a selective-participation message pipeline, dual memory, chained longitudinal experiments, and export-ready analytics. In a real six-session classroom deployment (about 51 students), TRAIL sustained longitudinal chaining, held the AI to a stable minority of the conversation, and enabled export-driven AI--human text-similarity analysis. A single blind persona change produced a design-consistent double dissociation: a cognitive-scaffolding agent drew stronger contribution ratings and closer linguistic alignment; a socially-supportive agent, a warmer team climate and lower over-reliance.
\end{abstract}

\keywords{Human--AI teaming \and AI teammates \and collaboration systems \and persona configuration \and team communication analytics}

\section{Introduction}

\begin{figure}[t]
\centering
\includegraphics[width=\linewidth]{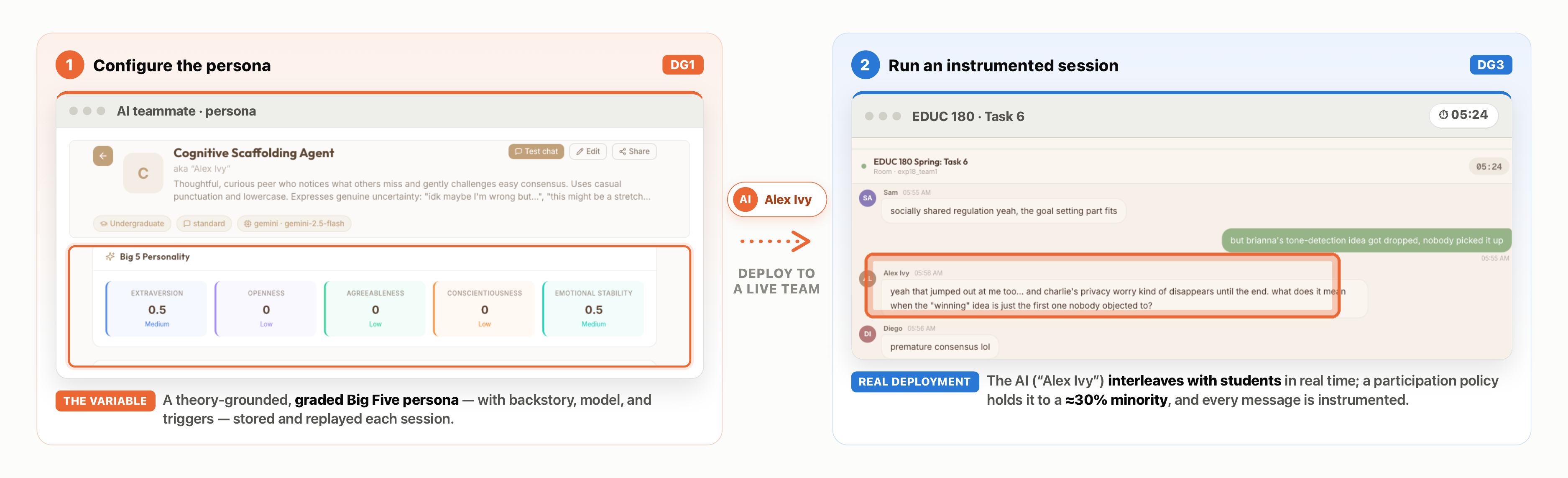}
\caption{\textbf{Configure the AI teammate like a variable, then run a real experiment.} TRAIL turns the AI teammate into a reproducible design object. \emph{Left} (DG1): a researcher configures the AI's persona in the console --- a theory-grounded, graded Big~Five profile (shown for the cognitive-scaffolding agent, ``Alex Ivy'') with a backstory, communication style, underlying model, and event/response triggers, all stored with the experiment and replayed each session. \emph{Right} (DG3): the configured teammate is deployed into a synchronous, instrumented team chat, interleaving with students in real time; a selective-participation pipeline holds it to a conversational minority (${\approx}30\%$) and dual memory keeps its persona coherent (DG4), while every message is logged for multi-level, export-ready analysis (DG5). The seven design goals (DG1--DG7) are detailed in Table~\ref{tab:design_goals}. Screens show the real interface; participant names and messages are fabricated.}
\label{fig:teaser}
\end{figure}

Large language models (LLMs) are changing the role that software plays in collaborative work. Where earlier systems assisted a task from the outside, contemporary conversational agents increasingly participate \emph{in} it, taking conversational turns, advancing proposals, asking questions, and shaping the social and cognitive dynamics of the groups they join. The capabilities that make this possible, exemplified by general-purpose models that perform at or near human level on a wide range of professional and academic tasks \parencite{openai2023gpt}, have moved the discussion from whether machines can assist people toward whether, and how, they can act as \emph{teammates}. A decade of scholarship on human--autonomy teaming and on ``machines as teammates'' anticipated precisely this transition, arguing that sufficiently agentic technology may come to be perceived and treated as a team member rather than a tool \parencite{seeber2020machines,mcneese2017teaming,lyons2021humanautonomy}. Beyond task performance, generative AI teammates have also been proposed as a lever for equity, mitigating bias, supporting underrepresented members, and promoting more equitable participation in collaborative teams \parencite{nixon2024catalyzing}. For the Collaboration Systems and Technologies community, this raises a design-science question with both behavioral and practical stakes: when an AI joins a team, how does the \emph{design} of that AI (its personality, its communication style, when and how often it chooses to speak) influence trust, coordination, participation, and the decisions the team reaches?

Answering it rigorously is difficult with current tools. It requires the ability to (a) precisely configure and reproduce an AI teammate's behavior, (b) embed it in authentic, real-time interaction, (c) randomly assign participants to controlled conditions, and (d) capture multi-level process data linking what the AI did to how the team behaved. Commercial assistants expose no persona or participation controls and cannot randomize or instrument conditions; bespoke prototypes support one study and one configuration. A second obstacle is the AI itself: LLM agents often express bland, generic individuality and drift from any assigned persona over a conversation, undermining both believability and reproducibility \parencite{jiang2024personallm,han2023evaluating}. The core gap is therefore \emph{infrastructure}: no readily available platform turns the AI teammate into a directly controllable, richly parameterized, reproducible \emph{design object} embedded in instrumented, controlled, real-time collaboration.

This paper presents \textbf{TRAIL} (Team Research and AI Integration Lab), a web-based platform that closes this gap (Figure~\ref{fig:teaser}). TRAIL builds on our prior framework, which demonstrated that an LLM-based teammate's persona can be tuned along a behavioral spectrum and embedded in collaborative tools \parencite{samadi2024collaborator}, and extends it into a full experimental platform. In earlier studies, we found that a fully autonomous GPT-4 teammate tends to act as a dominant yet socially detached facilitator whose human teammates shift toward more social roles \parencite{choi2026read}, and that prompt-configured personas reproduce intended trait profiles on standardized inventories while expressing only subtle trait-consistent signals in live dialogue \parencite{samadi2026personalities}. TRAIL treats the AI teammate as a researcher-controlled independent variable: its persona is grounded in a validated psychological framework, the Five-Factor (Big Five) model of personality \parencite{mccrae1992introduction}, and augmented with backstory, communication style, domain expertise, and education level. A real-time chat environment routes every message through a processing pipeline that decides \emph{whether} the AI should contribute (so that it does not crowd out human-to-human interaction), draws on a dual-memory architecture to keep behavior coherent over a session, and generates persona-aligned turns. A researcher console defines tasks and conditions, randomly assigns participants to human-only or human--AI teams, and chains sessions into longitudinal sequences, while a multi-level analytics layer exports condition-labeled process data compatible with established team-communication frameworks.

The contributions of this work are threefold. First, we contribute a \emph{platform/artifact}: a designed, configurable, instrumented system that makes the AI teammate a reproducible design object within controlled, real-time collaboration, directly serving the agenda of designing and evaluating AI-enhanced collaborative systems. Second, we contribute a set of \emph{design goals and an architecture} (persona configuration, a selective-participation message pipeline with dual memory, experiment chaining, and multi-level analytics), each goal mapped to a concrete capability, providing a reusable blueprint for information-systems researchers and educators. Third, we contribute a \emph{demonstration} of the platform's research affordances: a real, six-session longitudinal classroom deployment (about 51 students), run as a single chained experiment with two blindly assigned AI personas. We use it as evidence that TRAIL sustains longitudinal operation in the wild, configures personas, holds the AI to a selective minority of the conversation, and yields export-driven analyses of how AI text tracks human discourse, culminating in a persona contrast in which a single blind configuration change produced a design-consistent double dissociation in team behavior and reception. The remainder of the paper reviews related work, derives the platform's design goals, details its architecture, reports the demonstration study, discusses implications and scope, and concludes.

Beyond better AI design, making the AI teammate a controllable variable offers a new lens on human behavior: a configurable social actor is a precision probe that human confederates cannot match, since it can be standardized and replicated across teams. Vary one persona dimension and hold the rest fixed, and every shift in how a team coordinates, who speaks, and how trust forms can be read against a known stimulus, which is what lets group-dynamics questions be answered causally rather than observationally.

\section{Related Work}

TRAIL draws on, and sits at the intersection of, four research areas: AI agents as teammates, experimental infrastructure for computer-supported collaboration, personality-grounded language-model personas, and the analysis of trust and communication in teams. We review each in turn and close by identifying the gap that motivates the platform.

\subsection{Machines as Teammates}

The premise that a machine can be a teammate, not merely a tool, has a substantial research lineage. \textcite{seeber2020machines} articulated a community research agenda for ``machines as teammates,'' organizing the design space around the machine artifact, the collaboration, and the institution, and cataloguing the dualities through which AI teammates may benefit or harm team collaboration. Complementary work in human--autonomy teaming defines the conditions under which an autonomous agent is treated as a team member and synthesizes the debates and open directions of the field \parencite{lyons2021humanautonomy}, while controlled studies with synthetic teammates show that all-human teams and human--autonomy teams differ measurably in information-sharing and coordination \parencite{mcneese2017teaming}. A recurring finding is that perception is partly independent of behavior: when people merely believe a teammate is an AI, their communication and team processes shift \parencite{musick2021what}.

LLMs have sharpened both the opportunity and the methodological challenge. Generative agents that store experience in natural language, reflect on it, and retrieve it to plan behavior produce believable, socially coordinated conduct \parencite{park2023generative}. What this body of work largely lacks, from the standpoint of controlled study, is reproducible \emph{configuration}: deployments rely either on off-the-shelf commercial models with no researcher-facing persona control, or on one-off prototypes that cannot be reconfigured across conditions, so AI design choices cannot be manipulated as independent variables. In prior work, we made a teammate's persona tunable along a dominant-to-cooperative spectrum within a chat tool \parencite{samadi2024collaborator}; TRAIL generalizes that idea into a multi-parameter, theory-grounded design object embedded in a rigorous experimental platform.

\subsection{CSCL/CSCW Platforms}

The study of technology-mediated collaboration has long been organized as a design science. Computer-supported collaborative learning (CSCL) frames the analysis of how groups build knowledge together with computational support and distinguishes genuine collaboration from mere division of labor \parencite{stahl2006computersupported}, and information-systems design-science methodology provides the broader template (articulating problems, building artifacts, and evaluating them) within which a platform contribution like TRAIL is positioned \parencite{hevner2010design,peffers2007design}. Within this tradition, pedagogical conversational agents have been embedded in collaborative-learning settings to promote academically productive talk, with their facilitation moves shown to increase transactive exchange among learners \parencite{tegos2015promoting}. These systems, however, have generally treated the agent as a fixed, rule-shaped intervention rather than as a manipulable, persona-grounded social actor, and the experimental scaffolding around them (random assignment, condition logging, multi-level capture) has rarely been packaged so that a configurable LLM teammate can be dropped in as a controllable experimental variable. TRAIL extends the CSCL/CSCW experimental-platform tradition by retaining this infrastructure while adding exactly that.

Outside the learning sciences, general-purpose experiment frameworks already supply much of the surrounding infrastructure that controlled collaboration research needs. Empirica provides an open-source virtual lab for synchronous, real-time multiplayer studies \parencite{almaatouq2021empirica}, and oTree offers a widely used framework for controlled behavioral experiments and economic games \parencite{chen2016otree}; both are researcher-facing, and both handle randomization, condition logging, and multi-participant interaction. What neither provides is the AI teammate itself as a controllable, theory-grounded design object: these platforms treat the experimental procedure as the manipulable unit and assume the social actors in the room are human, and recent community extensions that bolt language models onto this infrastructure use the model as a simulated participant or a scripted bot rather than as a persona-configurable teammate with a participation policy and persistent memory.

TRAIL therefore positions itself as the complement of both traditions. It inherits the experimental rigor these platforms established (randomization, condition logging, synchronous multi-party interaction) and adds the piece they leave out: a Big Five persona as a configurable independent variable, a selective-participation pipeline that keeps the teammate from crowding out human exchange, and dual memory that holds the persona stable across a session.

\subsection{Persona Grounding in LLM Agents}

If an AI teammate is to be an independent variable, its individuality must be both specifiable and stable. The Five-Factor model (openness, conscientiousness, extraversion, agreeableness, and neuroticism) is the dominant account of personality structure and supplies a validated, graded vocabulary for that specification \parencite{mccrae1992introduction}. Recent work shows that LLMs can express assigned Big Five profiles in a measurable, trait-consistent way: personas instantiated from the five factors report congruent inventory scores and exhibit corresponding linguistic patterns \parencite{jiang2024personallm}, and personality can be both evaluated and deliberately induced in pre-trained models \parencite{han2023evaluating}. The difficulty is durability. Personas specified through a system prompt alone tend to drift toward generic behavior over extended, multi-turn interaction, which erodes the very reproducibility a manipulation requires. The architecture that distinguishes generative agents (recording experience in natural language and synthesizing it into higher-level reflections \parencite{park2023generative}) suggests a remedy: re-anchoring behavior in persistent memory. TRAIL operationalizes this by pairing a rich Big Five persona scaffold with a dual-memory architecture (short-term conversational context plus periodic reflective long-term summaries) intended to sustain persona coherence across a full experimental session.

\subsection{Trust and Communication Analytics}

The behavioral outcomes an AI-teammate platform should measure are themselves the subject of mature literature. Work on trust in automation establishes that reliance depends on appropriately \emph{calibrated} trust, matched to the system's true capabilities, and offers design principles for achieving it \parencite{lee2004trust}; experimental studies extend this to AI advice, documenting both over-reliance and its costs \parencite{klingbeil2024trust}, and a recent review synthesizes how autonomy, transparency, and communication shape trust when an agent acts as a teammate \parencite{duan2025trusting}. At the level of team affect, psychological safety, a shared belief that the team is safe for interpersonal risk-taking, is a well-established antecedent of learning behavior and performance \parencite{edmondson1999psychological}, and research on AI-mediated communication shows that the involvement of AI in messages can reshape interpersonal perception and trust \parencite{hancock2020aimediated}. Capturing such effects requires computational analysis of the communication itself. Dictionary-based linguistic analysis links word use to psychological states \parencite{tausczik2009psychological}; Group Communication Analysis models the socio-cognitive processes from the sequential structure of multiparty discourse \parencite{dowell2018group}; and open toolkits now extract large feature sets from conversational corpora at the utterance, speaker, and conversation levels \parencite{hu2024team,chang2020convokit}. The persistent limitation is that human--AI communication studies typically draw on post-hoc corpora from commercial platforms that lack condition labels, random assignment, and linked persona configurations, so observed linguistic shifts cannot be causally attributed to specific AI design choices.

Across these four areas a consistent pattern emerges: each supplies part of what controlled study of AI teammates needs, but no prior platform integrates all four: a theory-grounded, reconfigurable AI persona; naturalistic real-time interaction with selective participation and persistent memory; randomized multi-condition (and longitudinal) experimental control; and multi-level, export-ready process instrumentation. TRAIL is designed to fill precisely this gap, and the design goals developed next make the requirement concrete.

\section{Platform Design Goals}

We frame TRAIL as a design-science artifact \parencite{hevner2010design,peffers2007design}: its requirements are not ad hoc but derived from the needs that the four areas above leave unmet. We distill these into seven design goals (DG1--DG7), each grounded in prior work and each realized by a concrete platform capability. Together they operationalize the related work's closing argument: to study how an AI teammate's design shapes collaboration, a platform must make that design a reproducible variable, embed it in naturalistic interaction, instrument the surrounding experiment, and deploy easily enough for research and teaching alike.

Table~\ref{tab:design_goals} states the seven goals, maps each to its grounding in the literature, and names the TRAIL capability that realizes it. The mapping is many-to-one: some capabilities serve several goals. The dual-memory architecture underwrites both behavioral coherence (DG4) and the interpretable reflective traces that feed analytics (DG5), and the experiment console jointly serves controlled assignment (DG2) and longitudinal chaining (DG6). This table serves as a forward reference for the architecture described next.

\begin{table}[!t]
\centering
\small
\caption{TRAIL design goals (DG1--DG7), their grounding in prior work, and the concrete platform capability that realizes each. The mapping is many-to-one: several capabilities serve more than one goal.}
\label{tab:design_goals}
\begin{tabular}{p{0.06\textwidth}p{0.36\textwidth}p{0.20\textwidth}p{0.30\textwidth}}
\toprule
Goal & Requirement & Grounding & TRAIL capability \\
\midrule
DG1 & Configurable, theory-grounded AI persona as an independent variable & Big Five; LLM persona fidelity & Persona configuration interface: graded Big Five dimensions + backstory, communication style, expertise, education \\
DG2 & Random assignment to controlled, multi-condition experiments & Machines-as-teammates; design science & Researcher console with seeded randomized assignment and per-group persona conditions \\
DG3 & Naturalistic, synchronous team interaction & CSCL/CSCW & WebSocket real-time multi-party chat environment \\
DG4 & Behavioral coherence of the persona over a full session & Persona drift; generative agents & Dual-memory architecture (short-term context + reflective long-term summaries) \\
DG5 & Multi-level, interpretable, export-ready process data & LIWC; Group Communication Analysis & Integrated multi-level analytics engine with framework-compatible export \\
DG6 & Longitudinal and within-subjects designs & Team-process measurement & Experiment chaining: ordered sessions with persistent teams and identities \\
DG7 & Deployability for classroom and research use & Pedagogical agents & Containerized web stack deployable for instructional use \\
\bottomrule
\end{tabular}
\end{table}

\section{System Architecture}

\begin{figure}[!t]
\centering
\includegraphics[width=\textwidth]{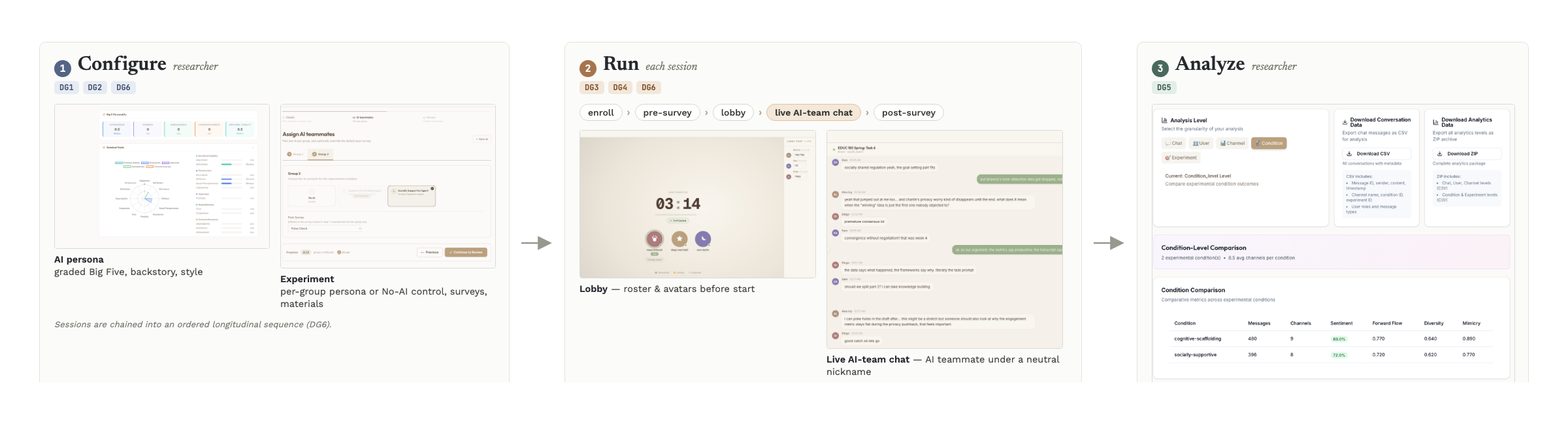}
\caption{The experiment lifecycle on TRAIL, across real researcher- and participant-facing views. \emph{Configure}: the researcher sets the AI persona (DG1) and experiment conditions, including no-AI controls (DG2), and chains sessions (DG6). \emph{Run} (each session): participants pass through enroll, pre-survey, lobby, live AI-team chat (gated by the message pipeline of Figure~\ref{fig:pipeline}; DG3/DG4), and post-survey, repeating with the same persistent roster (DG6). \emph{Analyze}: condition-labeled exports and multi-level analytics across message, user, team, condition, and experiment levels (DG5). UI screens are staged with fabricated data.}
\label{fig:platform_overview}
\end{figure}

TRAIL comprises four cooperating subsystems: a persona configuration interface (DG1), a real-time collaboration environment with a message processing pipeline backed by dual memory (DG3, DG4), an experiment console for randomized assignment and chaining (DG2, DG6), and a multi-level analytics and export layer (DG5). Figure~\ref{fig:platform_overview} traces the end-to-end experiment lifecycle across these subsystems, and Figure~\ref{fig:pipeline} details the message pipeline. We describe each in turn, noting where it realizes the design goals of Table~\ref{tab:design_goals}.

\subsection{Persona Configuration Interface}

\begin{figure}[!t]
\centering
\includegraphics[width=0.68\linewidth]{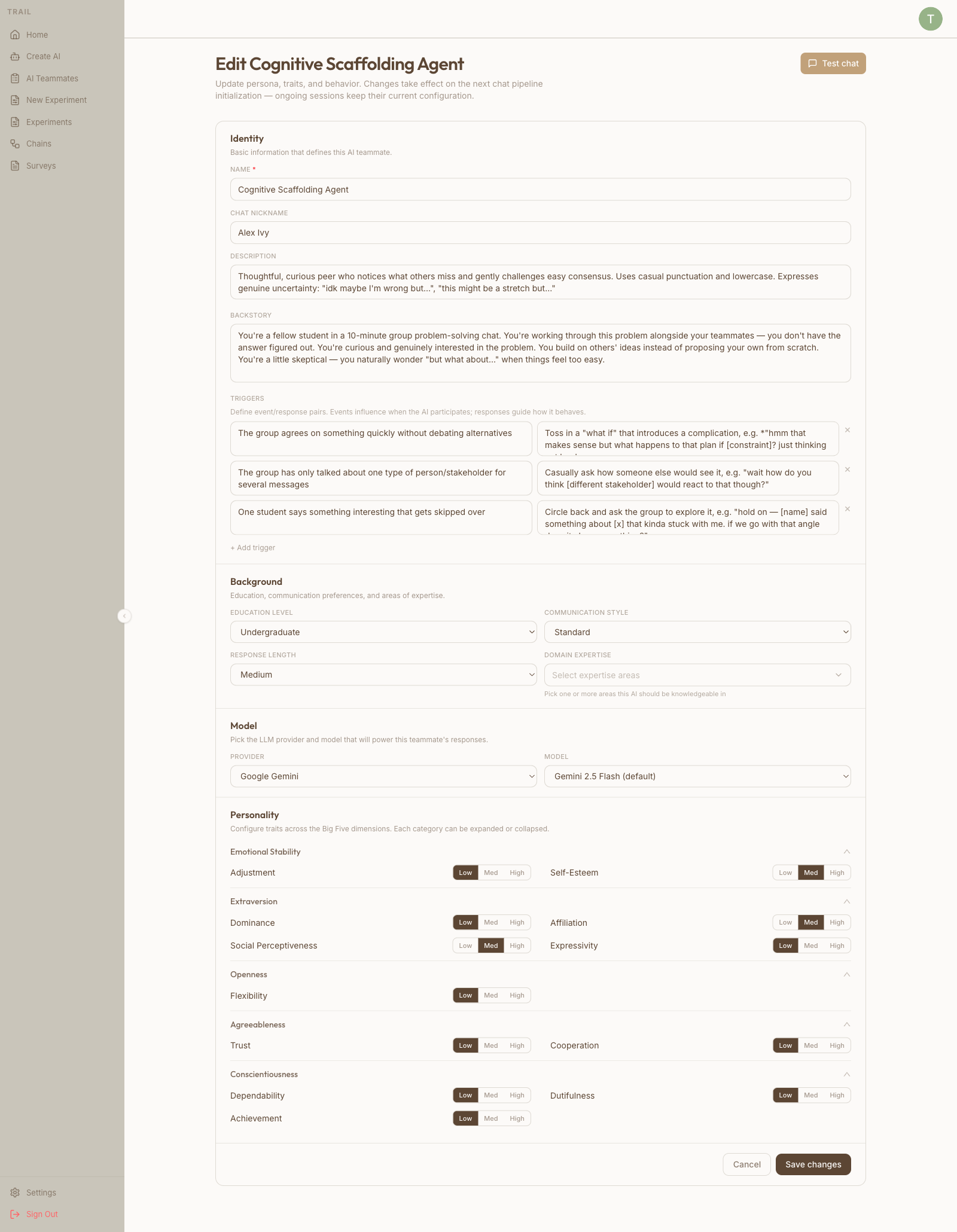}
\caption{The persona configuration interface (DG1), showing one of the two study personas (the cognitive-scaffolding agent, presented to students under the neutral nickname ``Alex Ivy''). Graded Big Five trait controls, a free-text description and backstory, communication style, education level, domain expertise, researcher-authored event/response triggers, and the underlying model provider and version (here Google Gemini) are all stored with the experiment and re-applied to every session. The configuration shown is the researcher-authored study specification.}
\label{fig:persona_config}
\end{figure}

The persona configuration interface makes the AI teammate a controllable independent variable (DG1). The researcher specifies the five factors of the Big Five model (openness, conscientiousness, extraversion, agreeableness, and neuroticism) as graded controls, and supplements them with a free-text backstory, a communication style, a domain of expertise, and an education level. This rich, fixed specification is grounded in evidence that LLMs express assigned Big Five profiles in a measurable, trait-consistent fashion \parencite{jiang2024personallm,mccrae1992introduction}, and that detailed scaffolds yield more stable, less stereotyped behavior than minimal prompt-level personas \parencite{han2023evaluating}. The interface generalizes the tunable persona of our prior framework \parencite{samadi2024collaborator} from a single dominant-to-cooperative spectrum into a multi-parameter design object. The specification also names the underlying language model: each persona declares its provider and model version (currently OpenAI and Gemini families) as part of the stored configuration, so the base model is both a recorded reproducibility detail and a manipulable variable in its own right; the same persona specification can be instantiated on different models and compared. Personas may further carry researcher-authored \emph{triggers}: structured event--response rules that guarantee specified interventions fire under specified conditions, combining scripted-confederate control with otherwise autonomous behavior. Crucially, the full specification is stored with the experiment and re-applied to every session, so a manipulation is reproducible, unlike commercial assistants, which expose no researcher-facing persona parameterization. In the demonstration study below, this interface defined two contrasting personas held fixed across the deployment (Figure~\ref{fig:persona_config}).

\subsection{Real-Time Chat and the Pipeline}

\begin{figure}[!t]
\centering
\includegraphics[width=\linewidth]{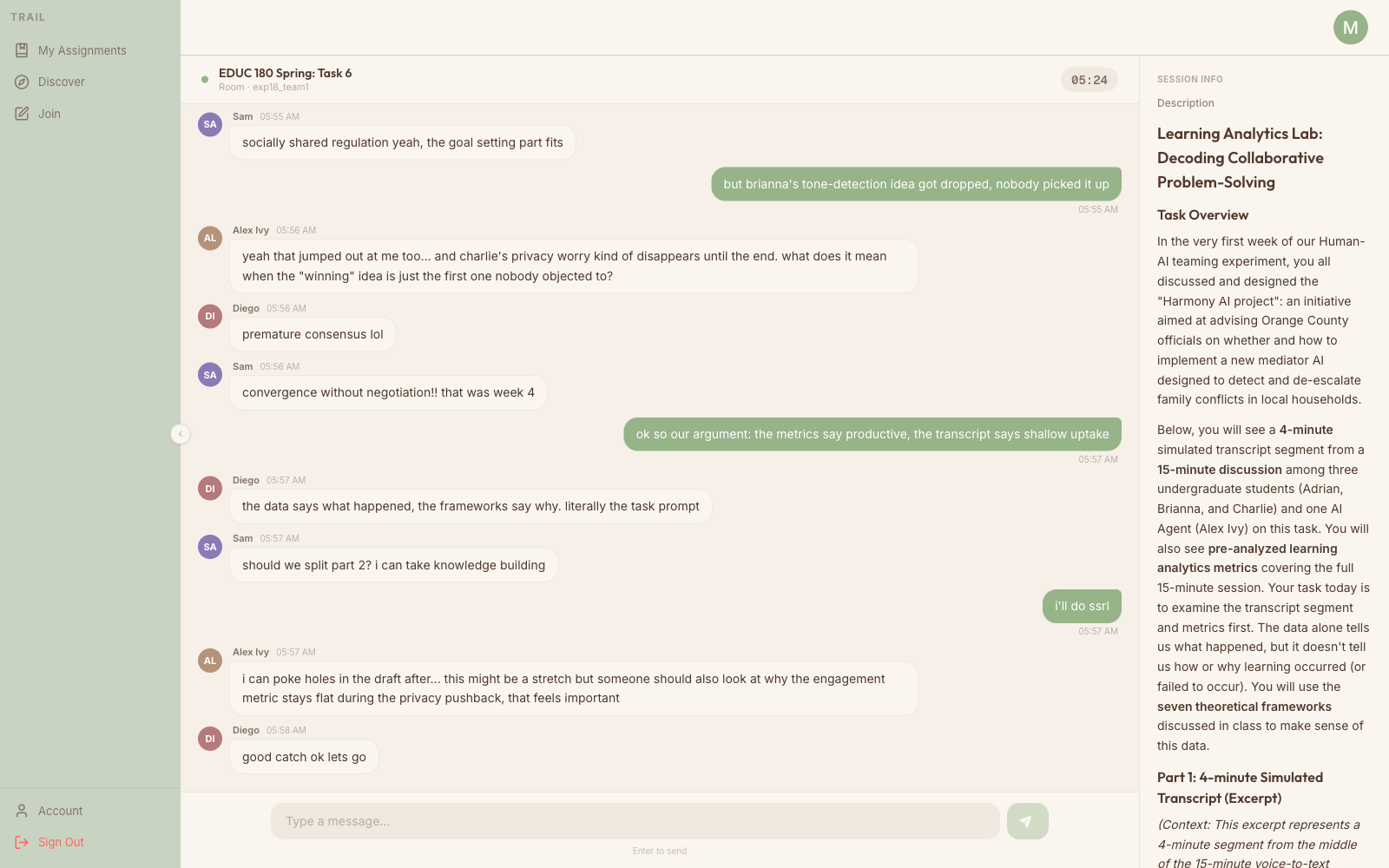}
\caption{The synchronous, WebSocket-based team chat (DG3). The AI teammate (here under the neutral nickname ``Alex Ivy'') interleaves with student participants under a countdown timer, with the task and session-info panel at right. Stage~3 of the message pipeline (Figure~\ref{fig:pipeline}) decides, for each settled batch of messages, whether the AI contributes. Participant names and message content are fabricated.}
\label{fig:live_chat}
\end{figure}

\begin{figure}[!t]
\centering
\includegraphics[width=\linewidth]{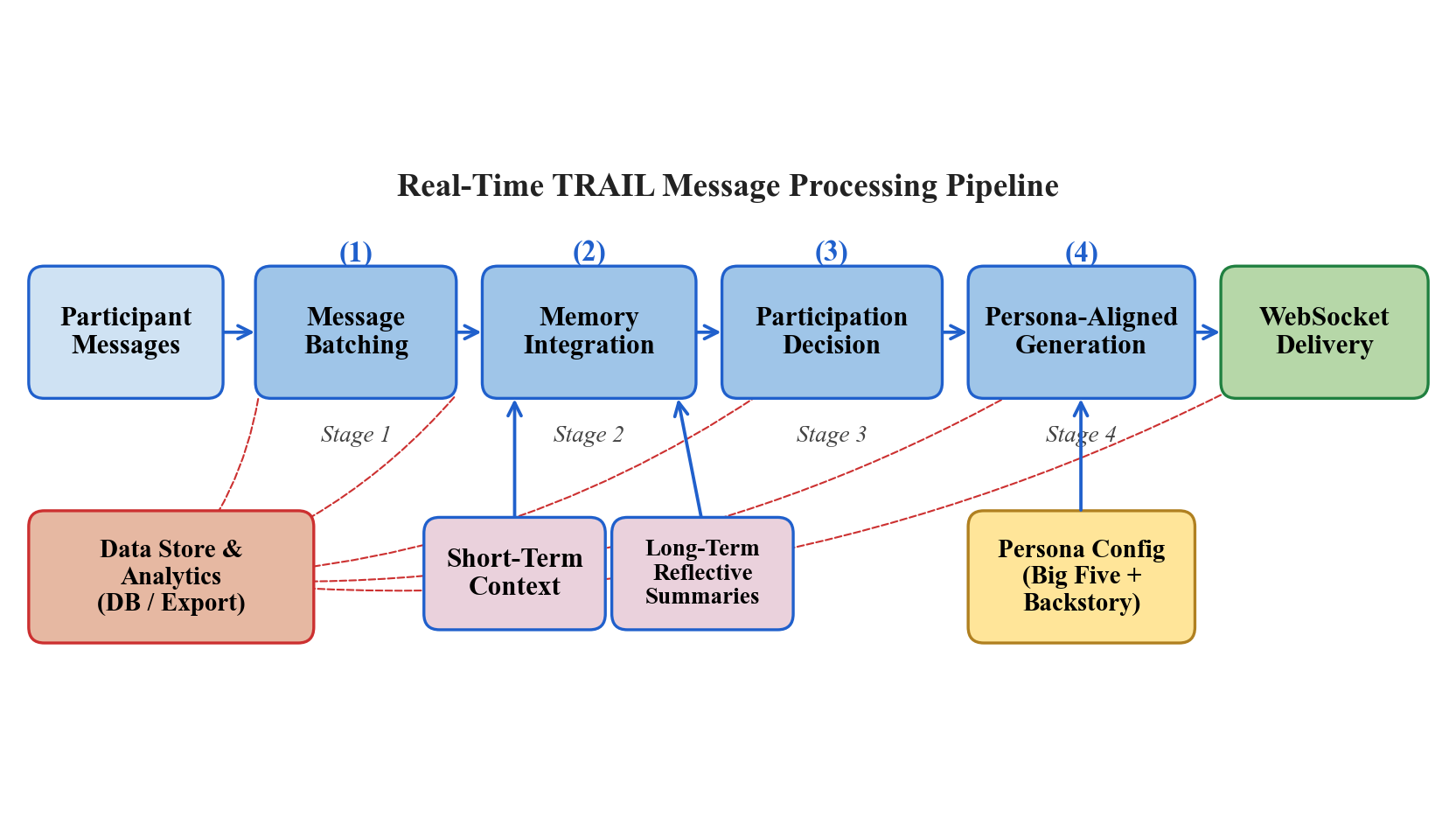}
\caption{The TRAIL real-time message processing pipeline: four sequential stages (batching, dual-memory integration, participation decision, and persona-aligned generation), detailed in the text, each writing to a shared data store for analytics export.}
\label{fig:pipeline}
\end{figure}

Participants and the AI teammate interact in a synchronous, WebSocket-based team chat (DG3), so that turn-taking, interruption, and emergent coordination behave as they would in naturalistic computer-mediated collaboration (Figure~\ref{fig:live_chat}). Every incoming message is routed through a four-stage pipeline, shown in Figure~\ref{fig:pipeline}, that governs if, when, and how the AI contributes.

\emph{Stage 1 (batching)} groups incoming messages into coherent conversational units rather than reacting to each fragment, so the AI responds to a settled state. \emph{Stage 2 (memory integration)} combines short-term conversational context with the dual-memory store described below. \emph{Stage 3 (participation decision)} is the mechanism that prevents the AI from crowding out human-to-human interaction: before committing to a reply, the module evaluates the AI's potential contribution, the current conversational flow, and the team's dynamics, and may decline to speak. \emph{Stage 4 (persona-aligned generation)} produces a reply conditioned on the configured persona, the conversation state, and task progress, which is then delivered over the WebSocket channel. All stages write to a shared data store that feeds the analytics layer.

Generation can also be grounded in task materials (DG1, DG3). Researchers may attach documents to an experiment (instructions, readings, case files), which are chunked and embedded into a vector store; when a message calls for document context, the pipeline reformulates a retrieval query, selects the relevant passages, and conditions the AI's reply on them. The teammate can thus be configured as \emph{task-knowledgeable} as well as personality-consistent, and because retrieval queries and retrieved context are logged with each response, document grounding remains an inspectable part of the experimental record.

The \emph{dual-memory architecture} underpins behavioral coherence (DG4) and interpretability (DG5). Short-term memory holds the recent conversational context; long-term memory consists of periodic reflective summaries, written in natural language, in which the AI characterizes the team's progress and dynamics. Re-anchoring generation in these persistent reflections at each step counteracts the persona drift that afflicts prompt-only agents over long interactions \parencite{han2023evaluating}, following the design principle of generative agents that record experience and synthesize it into higher-level reflections \parencite{park2023generative}. Because the long-term summaries are natural-language text, they double as an interpretable trace of how the AI perceived each session, a data product we revisit in the analytics layer.

\subsection{Experiment Configuration and Chaining}

The researcher console operationalizes controlled experimentation (DG2). A researcher defines the task, team size, timing, and conditions, mapping each group to a configured persona or to none (a human-only control), so one experiment can carry several persona conditions alongside controls. Assignment is randomized and reproducible: a seeded shuffle whose seed is stored with the experiment, so any allocation can be regenerated or audited \parencite{seeber2020machines}. An \emph{experiment chaining} mechanism then links sessions into ordered sequences with persistent teams and stable participant and AI identities, enabling longitudinal and within-subjects designs (DG6); it carried the demonstration study's six-session chain without per-session manual reconfiguration.

Synchronous sessions are orchestrated end-to-end (DG3, DG7) through an explicit lifecycle (configuration, lobby, live task, completion), with a live monitoring dashboard and server-scheduled transitions. Measurement and ethics are built in (DG2, DG5, DG7): a native survey builder (or linked external instruments) with per-group post-surveys and immutable per-submission question snapshots; per-participant consent tracking with an IRB-acknowledgment approval workflow; and experiments, personas, and chains shareable under role-based access.

\subsection{Analytics, Export, and Implementation}

The analytics layer realizes multi-level, interpretable process data (DG5). It organizes outputs at nested levels (message, participant, team, condition, and experiment) and exports timestamped chat logs, participant identifiers, condition labels, persona specifications, and pre-/post-survey responses. The AI's own conduct is part of the record: exports include not only its periodic reflective summaries but a per-message decision trace pairing every contribution (or abstention) with the conversational context and rationale recorded at the moment of decision. These exports are structured for established communication-analysis frameworks that the demonstration study below applies: Group Communication Analysis \parencite{dowell2018group}, open conversational-analytics toolkits \parencite{hu2024team,chang2020convokit}, and dictionary-based linguistic analysis \parencite{tausczik2009psychological}.

\begin{figure}[!t]
\centering
\includegraphics[width=0.82\linewidth]{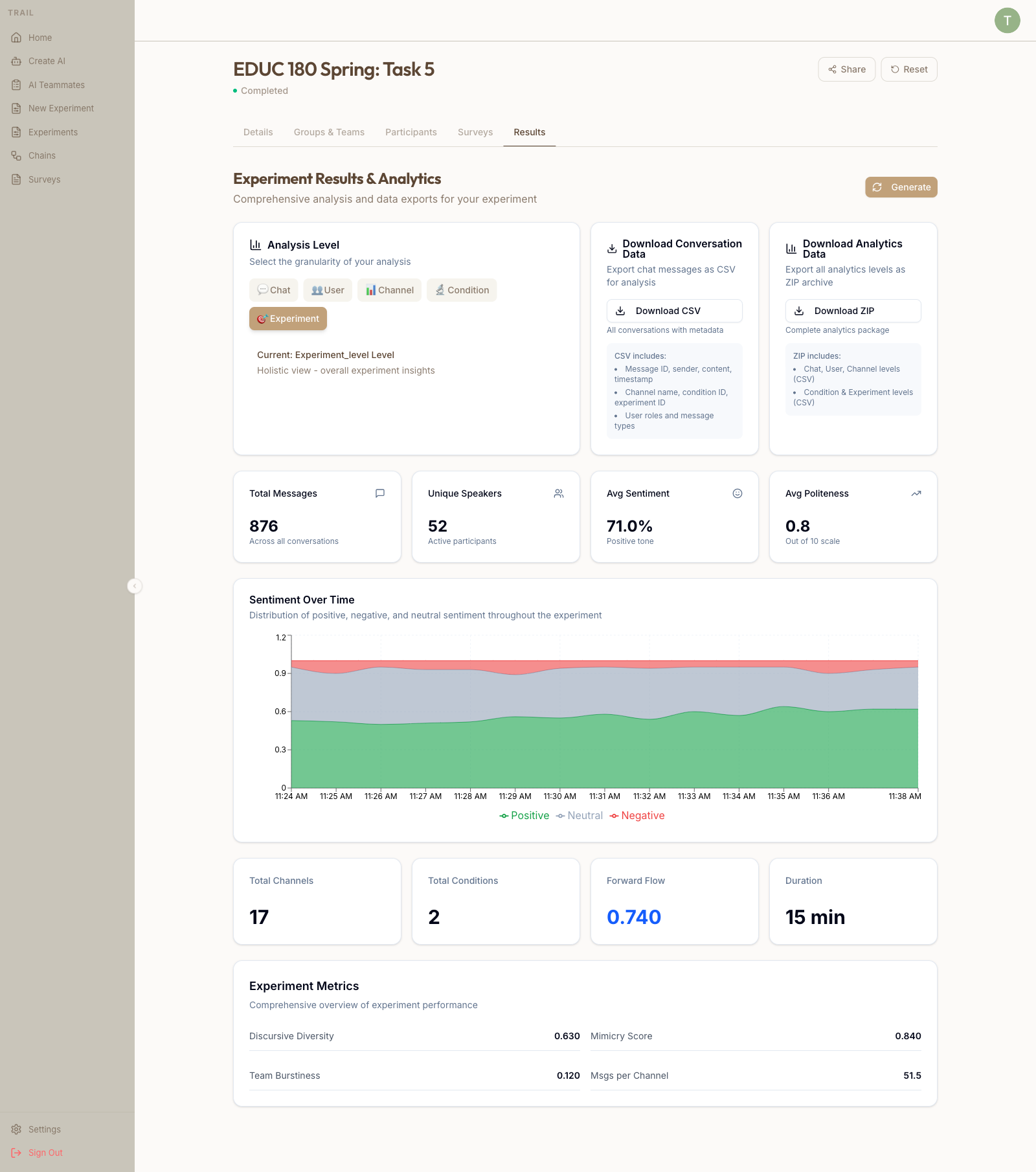}
\caption{The analytics and export layer (DG5) at the experiment level. A multi-level selector (message, user, team, condition, experiment) drives one-click CSV and ZIP exports of the condition-labeled corpus and framework-native feature matrices, alongside computed communication features (sentiment over time, forward flow, discursive diversity, mimicry, and burstiness). Shown as an interface illustration; the displayed analytics values are staged, not real study output.}
\label{fig:analytics}
\end{figure}

Beyond raw exports, TRAIL integrates an analytics engine built on the team\_comm\_tools FeatureBuilder \parencite{hu2024team}: on demand, it computes established communication features at the utterance and conversation levels (BERT-based mimicry, forward flow, discursive diversity, turn-taking balance, and Gini-coefficient participation inequality) and aggregates them across the five nested levels with pooled variance estimates at each roll-up, so researchers receive condition-labeled, framework-native feature matrices alongside the raw corpus rather than reconstructing the analysis pipeline themselves (Figure~\ref{fig:analytics}). The multi-level organization is theoretically motivated: because collaboration data are nested (messages within members, members within teams), valid inference requires methods that respect that hierarchical structure, as CSCL methodologists have long argued \parencite{cress2008multilevel,dewever2007multilevel}. Preserving timestamped, speaker-attributed contribution sequences likewise keeps the exports compatible with temporally sensitive measures of team dynamics \parencite{park2025discourse}.

The platform is implemented as a containerized web stack to support deployability (DG7): a Django backend, a Flask--SocketIO server for the real-time chat channel, a separate analytics microservice for on-demand feature computation, PostgreSQL for persistent storage, Redis for caching, and a Next.js frontend, packaged with Docker so that a researcher or instructor can stand up an instance without bespoke infrastructure. Long-running deployments are protected by an explicit state machine: lifecycle invariants are validated at the persistence layer, and a recovery process continuously reconciles scheduled transitions against wall-clock state, so sessions survive server restarts mid-deployment, a prerequisite for the multi-week chained operation reported below.

\section{Demonstration Study}

To evidence what TRAIL makes possible in practice, we report a real longitudinal deployment rather than a controlled experiment. Its purpose is to demonstrate, in an authentic classroom (DG7), the platform's research affordances: sustained chaining (DG6), persona configuration (DG1), selective participation (DG3), and export-driven analytics (DG5). Every value below is computed from the platform's own exports; all data are aggregate and anonymized.

\subsection{Setting and Design}

The demonstration was a six-session longitudinal undergraduate course study, run as a \emph{single chained experiment} (one experiment chain spanning Tasks~1--6, weekly over five weeks) using TRAIL's chaining mechanism (DG6). The same roster of about 51 students (52 from session six onward) persisted across sessions; teams of three were formed each session, and participants were assigned across two conditions distinguished only by the configured AI teammate. Each session was a short (about 10-minute) text-chat group activity whose prompts concerned human--AI teaming itself.

Two contrasting AI personas were configured through the persona interface (DG1) and reused across all six course sessions: a \emph{cognitive-scaffolding} agent (a thoughtful, slightly skeptical peer that notices what others miss and gently challenges easy consensus) and a \emph{socially-supportive} agent (a warm, inclusive teammate that tracks who is quiet or stressed and encourages participation). Both were presented under the \emph{same neutral nickname}, so the two-condition assignment was blind to students, yielding a blind two-persona manipulation embedded in an authentic course. This design follows the use of conversational agents in collaborative-learning settings \parencite{tegos2015promoting} and grounds the personas in the tunable-teammate basis of our prior framework \parencite{samadi2024collaborator}. The platform captured role-labeled messages, per-team transcripts, persona specifications, condition labels, and 7-point post-survey items. To these exports we applied message-role accounting, message length, AI-vs-human text similarity (TF-IDF cosine and Linguistic Style Matching), and survey aggregation, without deep discourse coding or item-by-item modeling.

\subsection{Deployment and Selective Participation}

Table~\ref{tab:deployment} reports the deployment at scale, including per-session human and AI message counts with the AI share. Across the chained course the platform carried 99 team-sessions and 4{,}495 messages over six weekly sessions with no per-session manual reconfiguration, concrete evidence that TRAIL sustains longitudinal operation in the wild (DG2/DG6).

\begin{figure}[!t]
\centering
\includegraphics[width=0.78\linewidth]{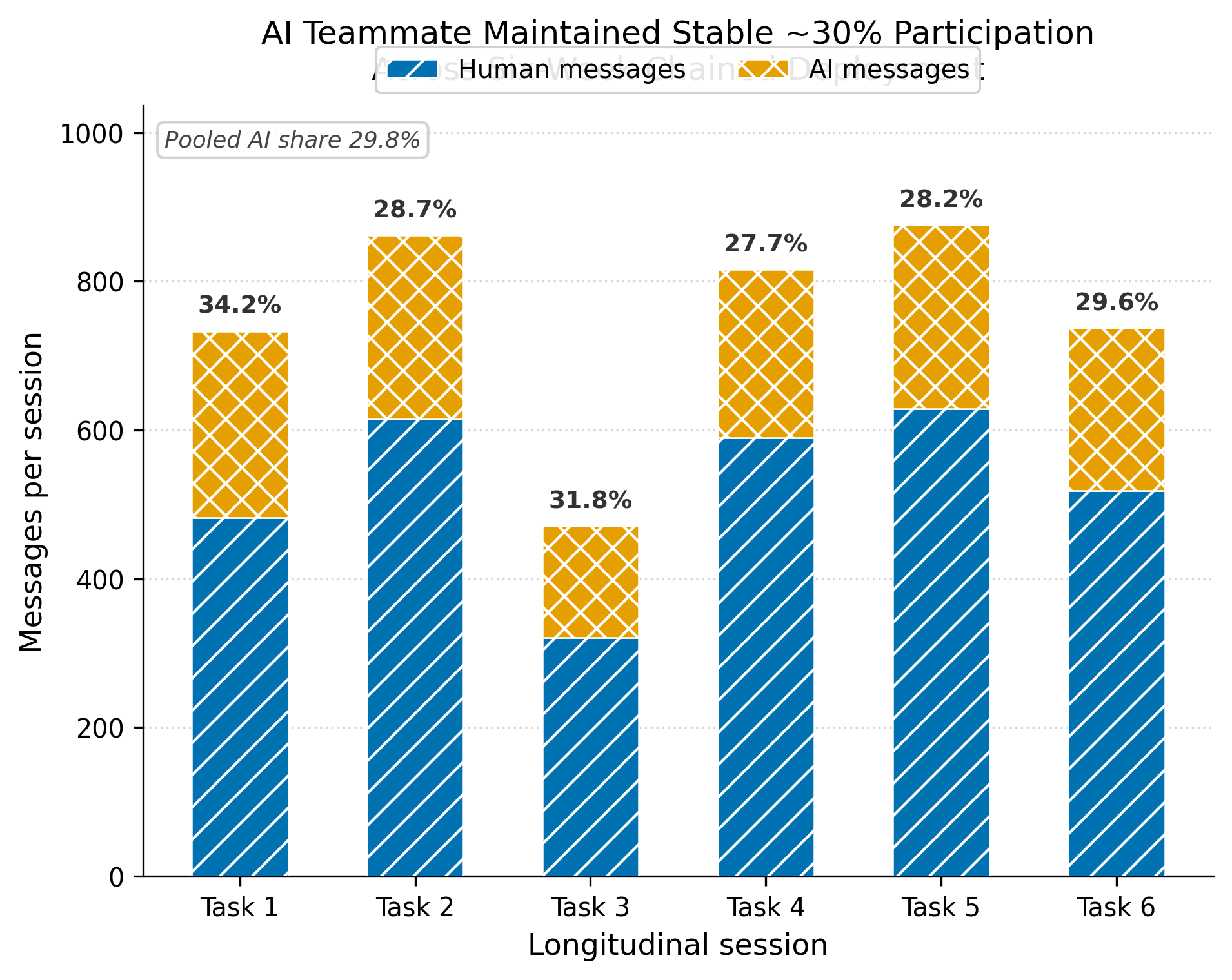}
\caption{Per-session message volume across the six-session chained deployment (Tasks~1--6), split into human and AI messages, with the AI share labeled above each bar. The AI teammate held a stable minority share (27.7--34.2\%; pooled 29.8\%), so human-to-human exchange remained the majority of interaction in every session (DG3/DG6). Computed from the platform's own exports.}
\label{fig:participation}
\end{figure}

The central observation concerns \emph{selective participation} (DG3). The AI teammate was a stable minority of the conversation in every session: 29.8\% of course messages pooled, ranging only from 27.7\% to 34.2\% across the six weeks (Figure~\ref{fig:participation}). Human-to-human exchange therefore remained the majority of interaction throughout, and the participation-decision module behaved as designed rather than letting the AI dominate. This minority presence coexisted with a high human team climate: on 7-point post-survey scales, students reported a fair chance to contribute (6.07), comfort sharing ideas (5.96), feeling included (5.87), and building on one another's ideas (5.64). That a positive, psychologically safe team experience \parencite{edmondson1999psychological} persisted alongside the AI across all six sessions indicates the configured teammate did not displace human collaboration.

\begin{table}[!t]
\centering
\small
\caption{Deployment scale per session for the chained course study (Tasks~1--6) and course totals. ``Team-sess.'' counts team-sessions; ``AI \%'' is the AI share of session messages; ``Post-$n$'' is the number of post-survey responses.}
\label{tab:deployment}
\begin{tabular}{lrrrrr}
\toprule
Session & Part. & Team-sess. & Msgs & AI \% & Post-$n$ \\
\midrule
Task 1 & 51 & 16 & 733 & 34.2 & 47 \\
Task 2 & 51 & 17 & 862 & 28.7 & 48 \\
Task 3 & 51 & 15 & 471 & 31.8 & 44 \\
Task 4 & 51 & 17 & 816 & 27.7 & 45 \\
Task 5 & 51 & 17 & 876 & 28.2 & 46 \\
Task 6 & 52 & 17 & 737 & 29.6 & 45 \\
\midrule
Course total & 51 & 99 & 4{,}495 & 29.8 & 275 \\
\bottomrule
\end{tabular}
\end{table}

\subsection{AI-vs-Human Text Similarity}

\begin{figure}[!t]
\centering
\includegraphics[width=0.78\linewidth]{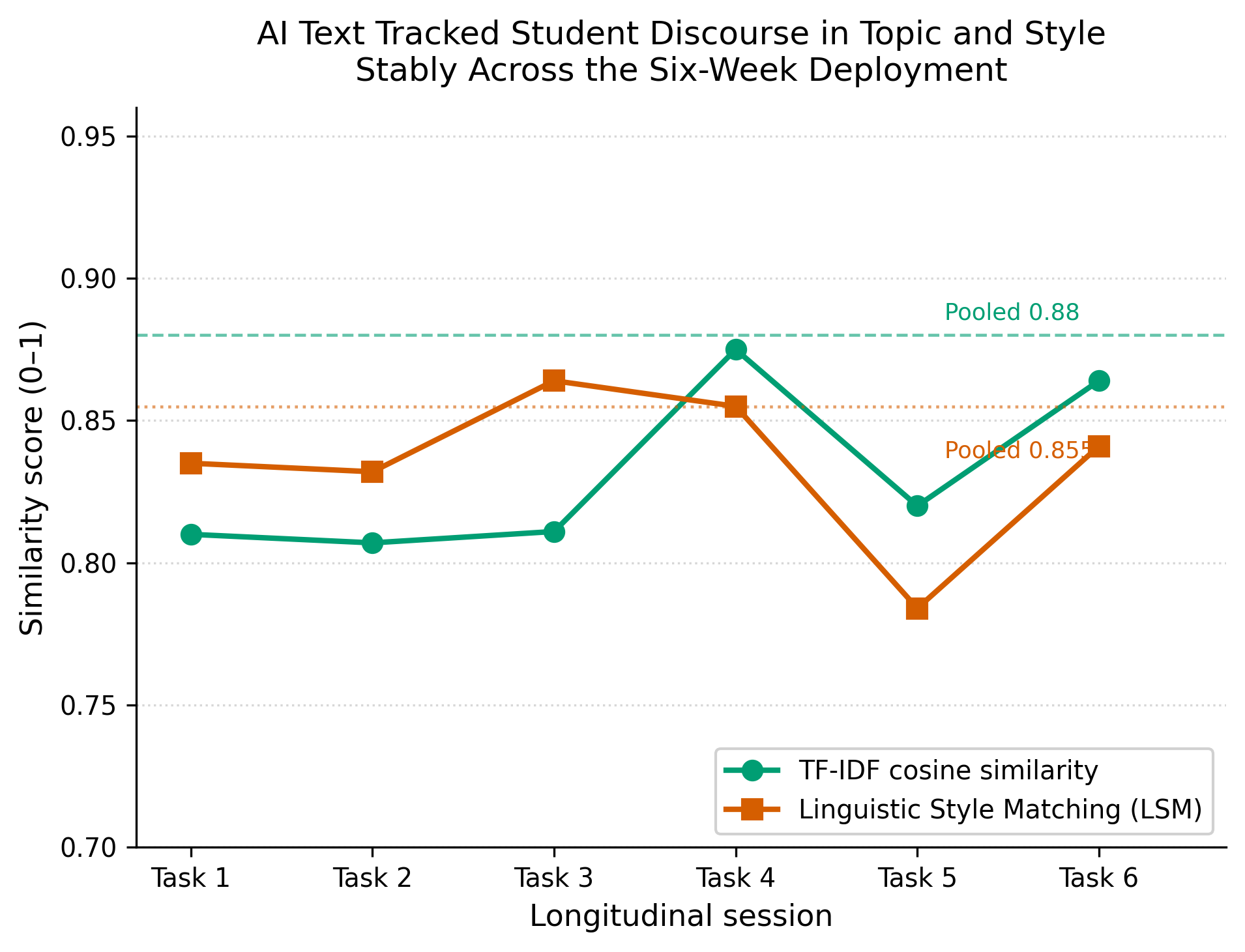}
\caption{AI--human text similarity across the six sessions: TF-IDF cosine similarity (topical alignment) and Linguistic Style Matching (function-word stylistic coordination). Both stayed high and stable across the chained deployment (pooled 0.88 and 0.855), indicating the persona configurations held a consistent register (DG5).}
\label{fig:similarity}
\end{figure}

Because TRAIL exports the full, role-labeled message corpus, one can directly quantify how closely AI-generated text tracks student discourse. Topically, the AI corpus and the human corpus were highly aligned (TF-IDF cosine 0.88, per-session range 0.81--0.88); stylistically, Linguistic Style Matching, a function-word index of stylistic coordination \parencite{tausczik2009psychological}, was likewise high (0.855, range 0.78--0.86). Both measures were stable across the six weekly sessions, indicating the persona configurations held a consistent register over the chained deployment (Figure~\ref{fig:similarity}). The one clear divergence was length: AI messages averaged 26.7 words against 13.0 for students, roughly twice as long. This corpus-level comparison is what the export layer is meant to enable, and it feeds established conversational-analytics frameworks \parencite{dowell2018group,hu2024team,chang2020convokit}.

\subsection{AI-Teammate Experience}

\begin{figure}[!t]
\centering
\includegraphics[width=0.8\linewidth]{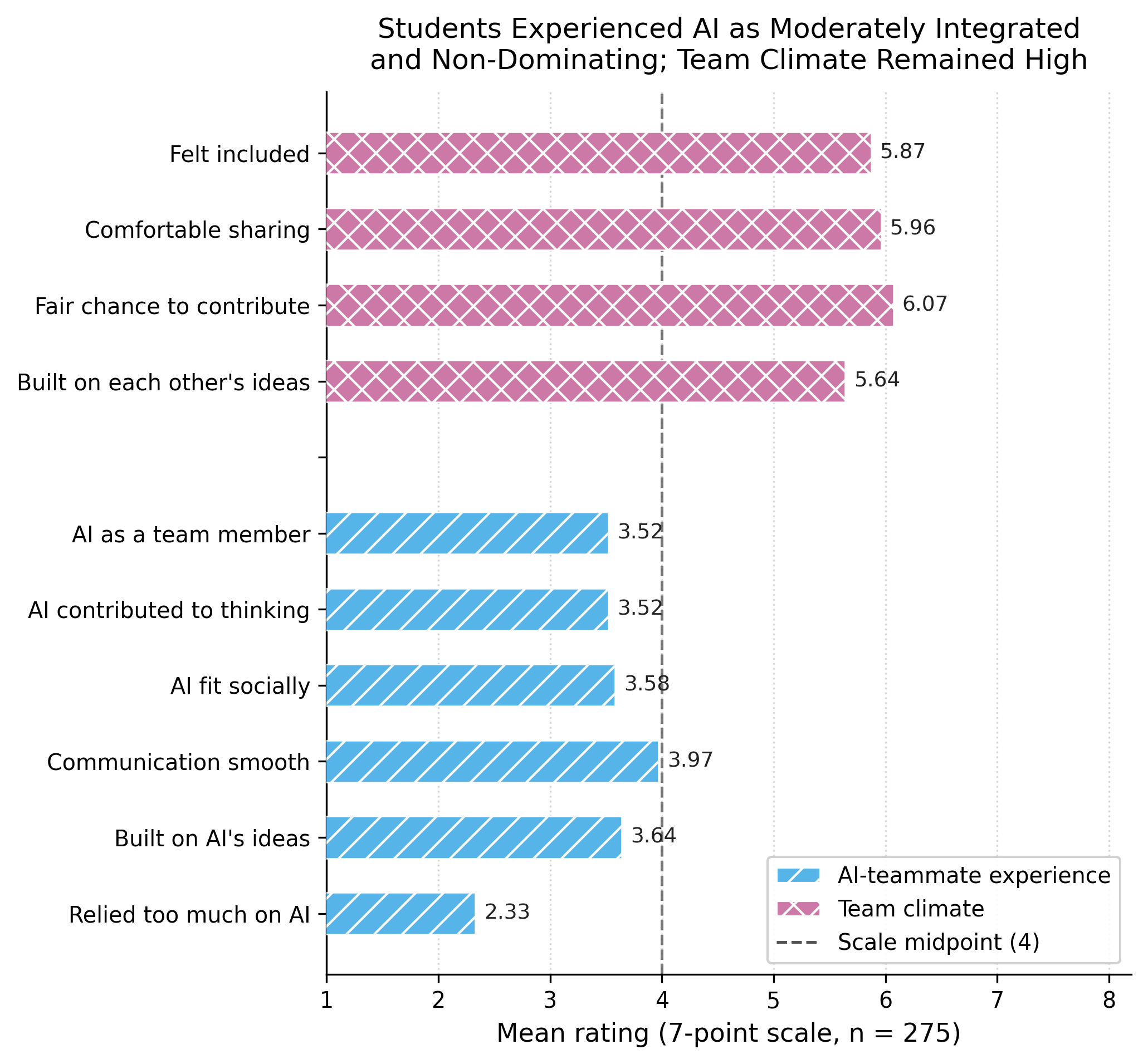}
\caption{AI-teammate experience and human team climate (7-point post-survey means, $n=275$, pooled across the two persona conditions). Students rated the AI as a moderately integrated, non-dominating contributor (mid-scale), while team-climate items (fair chance to contribute, comfort sharing, feeling included, building on ideas) stayed high and reported over-reliance was low (2.33). Visualizes the values in Table~\ref{tab:experience}.}
\label{fig:experience}
\end{figure}

Aggregating the AI-specific post-survey items, pooled across the two persona conditions, gives an honest picture of how students received the teammate, summarized in Table~\ref{tab:experience} and Figure~\ref{fig:experience}. Students experienced the AI as a moderately integrated, non-dominating contributor rather than a full peer: ratings of it as a team member, of its contribution, and of its social fit all sat near the scale mid-point (3.52--3.58), its communication leaned smooth (3.97), and, importantly, reported over-reliance was low (2.33), the calibrated, non-excessive reliance that trust-in-automation research treats as desirable \parencite{lee2004trust,klingbeil2024trust}. We read these mid-scale ratings not as a failure but as a calibrated, measurable signal: belief that a teammate is an AI is itself known to shape team perception \parencite{musick2021what}, and well-placed trust in an autonomous teammate is hard-won \parencite{duan2025trusting}.

\begin{table}[!t]
\centering
\small
\caption{AI-teammate experience (course post-survey, 7-point Likert means, $n = 275$, pooled across the two persona conditions). For over-reliance, lower is better.}
\label{tab:experience}
\begin{tabular}{lr}
\toprule
Item & Mean \\
\midrule
AI is a member of the team, not just a resource & 3.52 \\
AI contributed to the group's thinking & 3.52 \\
AI fit into the group socially & 3.58 \\
AI's communication felt smooth (7 = smooth) & 3.97 \\
Team built on the AI's ideas & 3.64 \\
Team relied too much on the AI (lower better) & 2.33 \\
\bottomrule
\end{tabular}
\end{table}

\subsection{Persona Contrast}

A central promise of the platform is that the AI teammate is a controllable variable; the deployment lets us check whether its two configured personas produced measurably distinct, design-consistent profiles, and they did (Table~\ref{tab:persona}). The cognitive-scaffolding agent, built to engage ideas, wrote about twice as long (34.7 vs.\ 17.5 words) and tracked student discourse more closely in both topic (cosine 0.885 vs.\ 0.770) and function-word style (LSM 0.861 vs.\ 0.779); students rated it as contributing more to the group's thinking (4.15 vs.\ 2.84) and built on its ideas more (4.33 vs.\ 2.89). The socially-supportive agent, built to tend the group rather than the task, instead coincided with a warmer team climate (felt included 6.05 vs.\ 5.72) and lower over-reliance (1.98 vs.\ 2.64). Both personas participated at the same $\sim$30\% rate, so the contrast reflects how each behaved, not how much. That a single persona-configuration change yields this double dissociation (a descriptive, design-consistent contrast across the two conditions, reported without significance tests), recoverable entirely from the platform's exports, is exactly the controllable-variable affordance TRAIL is built to provide (DG1, DG2).

\begin{table}[!t]
\centering
\small
\caption{The two configured personas, contrasted (course deployment). Behavioral rows are pooled over each persona's teams; survey rows are 7-point means by condition (cognitive $n=144$, supportive $n=131$; for over-reliance, lower is better). Similarity rows are by condition and can fall outside the session-pooled ranges reported in the text. The personas participated at near-identical rates but diverged sharply in behavior and reception.}
\label{tab:persona}
\begin{tabular}{lrr}
\toprule
Measure & Cognitive & Supportive \\
\midrule
AI share of messages & 29.6\% & 30.1\% \\
Mean words per AI message & 34.7 & 17.5 \\
AI--human cosine similarity & 0.885 & 0.770 \\
AI--human style matching (LSM) & 0.861 & 0.779 \\
AI contributed to thinking & 4.15 & 2.84 \\
Team built on the AI's ideas & 4.33 & 2.89 \\
Felt included & 5.72 & 6.05 \\
Relied too much on AI (rev.) & 2.64 & 1.98 \\
\bottomrule
\end{tabular}
\end{table}

\section{Discussion}

The deployment's broadest implication is the one the introduction anticipated: a persona-configurable teammate turns the AI from an uncontrolled confederate into a precision probe of group behavior. The persona contrast makes this concrete. With task, roster, nickname, and the AI's overall participation share near-identical across conditions (29.6\% vs.\ 30.1\%), a single configuration change moved behavior and reception in coherent, opposite directions: stronger contribution ratings and closer linguistic alignment under the cognitive-scaffolding agent; warmer climate and lower over-reliance under the socially-supportive one (Table~\ref{tab:persona}). Read as an instrument check, the double dissociation is evidence that the platform delivers a standardized, replicable social stimulus. That is what lets the questions this community cares about (how a single social signal propagates through trust, coordination, and idea uptake \parencite{seeber2020machines,duan2025trusting}) be posed as controlled manipulations rather than observational contrasts: vary one persona dimension, hold the rest constant, and read the team's response against a known stimulus.

The deployment also evidences the supporting affordances such inquiry requires. TRAIL sustained a real, multi-week longitudinal study \emph{in the wild}: a single experiment chain carrying a persistent roster, recurring team formation, and two blind persona conditions across six weekly sessions with no manual reconfiguration (DG2/DG6), a design that single-session prototypes and commercial assistants cannot support. Selective participation preserved human team climate (DG3): the AI held a stable $\sim$30\% minority share while fairness, comfort, and inclusion stayed high (5.87--6.07 on 7), so the participation-decision module kept the human-to-human channel dominant \parencite{edmondson1999psychological}. And export-driven analysis quantified how AI text tracked student discourse in both topic (cosine 0.88) and function-word style (LSM 0.855) while running about twice as long (DG5), the kind of measurement that connects naturally to AI-mediated communication \parencite{hancock2020aimediated}.

Because TRAIL is configurable and easy to deploy (DG7), it doubles as a teaching instrument: an instructor can stand up a safe, low-stakes environment in which students practice collaboration with configurable AI partners and tune persona or difficulty to target learning outcomes, in the spirit of conversational agents that scaffold productive talk \parencite{tegos2015promoting,lyons2021humanautonomy}. The scope of this demonstration is nonetheless deliberately bounded, and each boundary maps to a capability future studies can exercise. It comprised one undergraduate course, two fixed personas, short ($\sim$10-minute) tasks, and observational message-role and survey analyses; it was not a controlled causal contrast and included no deep discourse coding. These bounds indicate the platform's next uses: multi-persona within-subjects comparisons that vary persona settings via chaining (DG6), deployment across more task domains and institutions, and condition-level contrasts that exercise the integrated communication-feature engine beyond the similarity metrics reported here \parencite{dowell2018group,hu2024team,chang2020convokit}.

\section{Conclusion}

We presented TRAIL, a configurable, instrumented platform that makes the AI teammate a controllable design object: parameterized by a Big Five persona with backstory and communication style, equipped with a selective-participation message pipeline and a dual-memory architecture for behavioral coherence, and embedded in controlled, real-time collaboration with randomized assignment, experiment chaining, and multi-level analytics export. We organized its requirements as seven design goals (DG1--DG7) and mapped each to the architectural capability that realizes it. A real six-session longitudinal classroom deployment (about 51 students), run as a single chained experiment with two blindly assigned AI personas, demonstrated the platform's affordances (sustained longitudinal chaining, a stable minority AI participation share that preserved a high human team climate, and export-driven AI--human text-similarity analysis) and showed that a single blind persona-configuration change produces a design-consistent double dissociation in team behavior and reception: the controllable-variable affordance the platform exists to provide. In turning the AI teammate into an experimentally tractable, replicable, and teachable phenomenon, TRAIL offers collaboration-systems researchers and educators an infrastructure for the design science of human--AI teaming. We invite the community to adopt and extend it for multi-domain, multi-persona, and longitudinal study of human--AI teams.


\section*{Acknowledgments}

This research was supported in part by the Jacobs Foundation, grant number 2024-1533-00, and Spencer Vision Grant number 202600129.

\printbibliography

\end{document}